\documentclass[twocolumn,pra,superscriptaddress]{revtex4-2}
\usepackage{mathtools,amsfonts,amsmath,amssymb,mathrsfs}
\usepackage{empheq}
\usepackage{calrsfs}
\usepackage{exscale}
\usepackage{xfrac}
\usepackage{stmaryrd}
\usepackage{eucal}
\usepackage{graphicx}
\usepackage{epsfig} 
\usepackage{graphics}
\usepackage{xcolor}
\usepackage{color}    
\usepackage{bm}
\usepackage{float}
\usepackage{hyperref} 
\usepackage{multirow}  
\hypersetup{colorlinks,linkcolor={blue},citecolor={red},urlcolor={red}} 
\graphicspath{{../mygraphics/}}

\begin{document}

\bibliographystyle{apsrev}
%
%
\title{ Rabi oscillations, entanglement and 
        teleportation in the anti-Jaynes-Cummings 
        model
      }
%
\author{Christopher Mayero}
\email[E-mail address: ]{cmayero@yahoo.com}
\author{Joseph Akeyo Omolo}   
\author{Onyango Stephen Onyango}
\affiliation{
Maseno University, Department of Physics and Materials Science, Private Bag-40105, Maseno, Kenya.}
\date{January 20, 2021}
%
%
%
\begin{abstract}
This paper provides a scheme for generating maximally entangled qubit states in the anti-Jaynes-Cummings interaction mechanism, so called entangled anti-polariton qubit states. We demonstrate that in an initial vacuum-field, Rabi oscillations in a cavity mode in the anti-Jaynes-Cummings interaction process, occur in the reverse sense relative to the Jaynes-Cummings interaction process and that time evolution of entanglement in the anti-Jaynes-Cummings interaction process takes the same form as in the Jaynes-Cummings interaction process. With the generated anti-polariton qubit state as one of the initial qubits, we present quantum teleportation of an atomic quantum state by applying entanglement swapping protocol achieving an impressive maximal teleportation fidelity~$F_\rho=1$. \\

\noindent
\textbf{Keywords}: Jaynes-Cummings, anti-Jaynes-Cummings, Rabi oscillations, entanglement, entanglement swapping, teleportation, maximal teleportation fidelity  
\end{abstract}

\maketitle
%
%
\section{Introduction}
\label{sec:intro}
The basic model of quantized light-matter interaction describing a two-level atom coupled to a single mode of quantized electromagnetic radiation is the quantum Rabi model \cite{braak2011integrability, omolo2017conserved, omolo2017polariton,omolo2019photospins}. Recently, it has been shown that the operator ordering principle distinguishes the Jaynes-Cummings (JC) and anti-Jaynes-Cummings (AJC) Hamiltonians \cite{omolo2017conserved, omolo2017polariton, omolo2019photospins} as normal and anti-normal order components of the quantum Rabi model. In this approach the JC interaction represents the coupling of a two-level atom to the rotating positive frequency component of the field mode while the AJC interaction represents the coupling of the two-level atom to the anti-rotating (anti-clockwise) negative frequency component of the field mode,  because the electromagnetic field mode is composed of positive and negative frequency components \cite{born1999principles}. 

The long-standing challenge of determining a conserved excitation number and  corresponding $U(1)$ symmetry operators for the AJC component was finally solved in \cite{omolo2017conserved}. The discovery and proof of a conserved excitation number operator of the AJC Hamiltonian \cite{omolo2017conserved} now means that dynamics generated by the AJC Hamiltonian is exactly solvable, as demonstrated in the polariton and anti-polariton qubit (photospin qubit) models in \cite{omolo2017polariton, omolo2019photospins}. 

Noting that the JC model has been extensively studied in both theory and experiment in quantum optics, we now focus attention on the AJC model which has not received much attention over the years due to the erroneously assumed lack of a conserved excitation number operator. The reformulation developed in  \cite{omolo2017conserved, omolo2017polariton, omolo2019photospins}, drastically simplifies exact solutions of the AJC model, which we shall here apply.

In this paper, we are interested in analysis of quantum state configuration of the qubit states, entanglement of qubits in the AJC model and the application of the entangled qubit state vectors in teleportation of an entangled atomic quantum state. The content of this paper is therefore summarized as follows. Section \ref{sec:model} presents an overview of the theoretical model. In section \ref{sec:rabi},  Rabi oscillations in the AJC model is studied. In section \ref{sec:entquant}, entanglement of AJC qubit state vectors is analysed. In section \ref{sec:teleportation}, teleportation as an application of entanglement is presented and finally section \ref{sec:conclusion} presents the conclusion. 
%
%
\section{The model}
\label{sec:model}
The quantum Rabi model of a quantized electromagnetic 
field mode interacting with a two-level atom is 
generated by the Hamiltonian \cite{omolo2017conserved} 
\begin{equation}
\hat{H}_R=\frac{1}{2}\hbar\omega(\hat{a}^{\dagger}
\hat{a}+\hat{a}\hat{a}^{\dagger})+
\hbar\omega_0\hat{s}_z + \hbar\lambda(\hat{a}
+\hat{a}^{\dagger})(\hat{s}_++\hat{s}_-)
\label{eq:rabi1}
\end{equation}
noting that the free field mode Hamiltonian is expressed in normal and anti-normal order form $\frac{1}{2}\hbar\omega(\hat{a}^{\dagger}\hat{a}+\hat{a}\hat{a}^{\dagger})$.
Here, $\omega\hspace{1mm},\hspace{1mm}\hat{a}\hspace{1mm},\hspace{1mm}\hat{a}^{\dagger}$ are quantized field mode angular frequency, annihilation and creation operators, while $\omega_0,\,\hat{s}_z\,\hat{s}_+,\,\hat{s}_-$ are atomic state transition angular frequency and operators. The Rabi Hamiltonian in Eq.~\eqref{eq:rabi1} is expressed in a symmetrized two-component form \cite{omolo2017conserved, omolo2017polariton, omolo2019photospins}
\begin{equation}
\hat{H}_R=\frac{1}{2}(\hat{H}+\hat{\overline{H}}\,)
\label{eq:rabi2}
\end{equation}
where $\hat{H}$ is the standard JC Hamiltonian 
interpreted as a polariton qubit Hamiltonian 
expressed in the form \cite{omolo2017conserved}
\begin{eqnarray}
\hat{H} &=&\hbar\omega\hat{N}+2\hbar\lambda\hat{A}-\frac{1}{2}\hbar\omega 
\quad ; \quad 
\hat{N} =\hat{a}^{\dagger}\hat{a}+\hat{s}_+\hat{s}_- 
\nonumber \\ 
\hat{A} &=&\alpha\hat{s}_z+\hat{a}\hat{s}_++\hat{a}^{\dagger}\hat{s}_-
\quad ; \quad 
\alpha = \frac{\omega_0-\omega}{2\lambda}
\label{eq:pham1}
\end{eqnarray}
while $\hat{\overline{H}}$ is the AJC Hamiltonian interpreted as an anti-polariton qubit Hamiltonian in the form \cite{omolo2017conserved}
\begin{eqnarray}
\hat{\overline{H}} &=& \hbar\omega\hat{\overline{N}}+2\hbar\lambda\hat{\overline{A}}-\frac{1}{2}\hbar\omega
\quad ; \quad
\hat{\overline{N}}=\hat{a}\hat{a}^{\dagger}+\hat{s}_-\hat{s}_+
\nonumber \\ 
\hat{\overline{A}}&=&\overline{\alpha}\hat{s}_z+\hat{a}\hat{s}_-+\hat{a}^{\dagger}\hat{s}_+
\quad ; \quad
\overline{\alpha}=\frac{\omega_0+\omega}{2\lambda}\;.
\label{eq:antpham1}
\end{eqnarray}
In Eqs.~\eqref{eq:pham1} and \eqref{eq:antpham1}, $\hat{N}$, $\hat{\overline{N}}$ and $\hat{A}$, $\hat{\overline{A}}$ are the respective polariton and anti-polariton qubit conserved excitation numbers and state transition operators.

Following the physical property established in  \cite{omolo2019photospins}, that for the field mode in an initial vacuum state only an atom in an initial excited state $|e\rangle$ entering the cavity  couples to the rotating positive frequency field  component in the JC interaction mechanism, while only an atom in an initial ground state $|g\rangle$ entering the cavity couples to the anti-rotating negative frequency field  component in an AJC interaction mechanism, we generally take the atom to be in an initial excited state $|e\rangle$ in the JC model and in an initial ground state $|g\rangle$ in the AJC model. 

Considering the AJC dynamics, applying the state transition operator $\hat{\overline{A}}$ from Eq.~\eqref{eq:antpham1} to the initial atom-field \textit{n}-photon ground state vector $|g,n\rangle$, the basic qubit state vectors $|\psi_{gn}\rangle$ and $|\overline{\phi}_{gn}\rangle$ are determined in the form (\textit{n}=0,1,2,....) \cite{omolo2019photospins}
\begin{equation}
|\psi_{gn}\rangle=|g,n\rangle \quad ;\quad 
|\overline{\phi}_{gn}\rangle=-\overline{c}_{gn}|g,n\rangle+\overline{s}_{gn}|e,n+1\rangle
\label{eq:entsuptate}
\end{equation}
with dimensionless interaction parameters $\overline{c}_{gn}$, $\overline{s}_{gn}$ and Rabi frequency $\overline{R}_{gn}$ defined as
\begin{eqnarray}
\overline{c}_{gn} &=& \frac{\overline{\delta}}{2\overline{R}_{gn}}
\quad ;\quad
\overline{s}_{gn}=\frac{2\lambda\sqrt{n+1}}{\overline{R}_{gn}}
\quad; \quad 
\overline{R}_{gn}=2\lambda{\overline{A}_{gn}}
\nonumber \\
\overline{A}_{gn} &=&\sqrt{(n+1)+\frac{\overline{\delta}^2}{16\lambda^2}}
\quad ; \quad
\overline{\delta}=\omega_0+\omega
\label{eq:parameters}
\end{eqnarray}
where we have introduced sum frequency $\overline{\delta}=\omega_0+\omega$ to redefine $\overline{\alpha}$ 
in Eq.~\eqref{eq:antpham1}.

The qubit state vectors in Eq.~\eqref{eq:entsuptate} satisfy the qubit state transition algebraic operations
\begin{equation}
\hat{\overline{A}}|\psi_{gn}\rangle=\overline{A}_{gn}|\overline{\phi}_{gn}\rangle
\quad ; \quad
\hat{\overline{A}}|\overline{\phi}_{gn}\rangle=\overline{A}_{gn}|\psi_{gn}\rangle
\label{eq:traans}
\end{equation}
In the AJC qubit subspace spanned by normalized but non-orthogonal basic qubit state vectors  
$|\psi_{gn}\rangle$, $|\overline{\phi}_{gn}\rangle$  the basic qubit state transition operator $\hat{\overline{\varepsilon}}_g$ and identity operator $\hat{\overline{I}}_g$ are introduced according to the definitions \cite{omolo2019photospins}

\begin{equation}
\hat{\overline{\varepsilon}}_g=\frac{\hat{\overline{A}}}{\overline{A}_{gn}}
\quad ; \quad
\hat{\overline{I}}_g=\frac{\hat{\overline{A}}^2}{\overline{A}_{gn}^2}\quad \Rightarrow\quad \hat{\overline{I}}_g=\hat{\overline{\varepsilon}}_g^2
\label{eq:anttransop1}
\end{equation}
which on substituting into Eq.~\eqref{eq:traans} generates the basic qubit state transition algebraic operations
\begin{eqnarray}
\hat{\overline{\varepsilon}}_g|\psi_{gn}\rangle &=& 
|\overline{\phi}_{gn}\rangle
\quad ; \quad
\hat{\overline{\varepsilon}}_g|\overline{\phi}_{gn}
\rangle=|\psi_{gn}\rangle
\nonumber \\
\hat{\overline{I}}_g|\psi_{gn}\rangle &=&|\psi_{gn}\rangle
\quad ; \quad
\hat{\overline{I}}_g|\overline{\phi}_{gn}\rangle = 
|\overline{\phi}_{gn}\rangle
\label{eq:algop11}
\end{eqnarray}
The algebraic properties 
$\hat{\overline{\varepsilon}}_g^{2k}=\hat{\overline{I}}_g$ 
and $\hat{\overline{\varepsilon}}_g^{2k+1}=\hat{\overline{\varepsilon}}_g$ 
easily gives the final property \cite{omolo2019photospins}
\begin{equation}
e^{-i\theta\hat{\overline{\varepsilon}}_g} 
=\cos(\theta)\hat{\overline{I}}_g-i\sin(\theta)\hat{\overline{\varepsilon}}_g
\label{eq:antialgprop}
\end{equation}
which is useful in evaluating time-evolution operators.

The AJC qubit Hamiltonian defined within the qubit 
subspace spanned by the basic qubit state vectors 
$|\psi_{gn}\rangle$ , $|\overline{\phi}_{gn}\rangle$ is 
then expressed in terms of the basic qubit states 
transition operators 
$\hat{\overline{\varepsilon}}_g$, $\hat{\overline{I}}_g$ 
in the form \cite{omolo2019photospins}
\begin{equation}
\hat{\overline{H}}_g=\hbar\omega(n+\frac{3}{2})\hat{\overline{I}}_g+\hbar\overline{R}_{gn}\hat{\overline{\varepsilon}}_g \,.
\label{eq:antijch2}
\end{equation}
We use this form of the AJC Hamiltonian to determine 
the general time-evolving state vector describing Rabi 
oscillations in the AJC dynamics in Sec.~\ref{sec:rabi} below.

%
%
\section{Rabi oscillations}
\label{sec:rabi}

The general dynamics generated by the AJC Hamiltonian in Eq.~\eqref{eq:antijch2} is described by a time evolving AJC qubit state vector 
$\displaystyle |\overline{\Psi}_{gn}(t)\rangle$ obtained from the time-dependent Schr\"odinger equation in the form \cite{omolo2019photospins}
\begin{equation}
|\overline{\Psi}_{gn}(t)\rangle=\hat{\overline{U}}_g(t)|\psi_{gn}\rangle 
\quad ;\quad
\hat{\overline{U}}_g(t)=e^{-\frac{i}{\hbar}\hat{\overline{H}}_gt}
\label{eq:schtevol}
\end{equation}
where $\hat{\overline{U}}_g(t)$ is the time evolution operator. Substituting $\hat{\overline{H}}_g$ from Eq.~\eqref{eq:antijch2} into Eq.~\eqref{eq:schtevol} and applying appropriate algebraic properties \cite{omolo2019photospins}, we use the relation in Eq.~\eqref{eq:antialgprop} to express the time evolution operator in its final form 
\begin{equation}
\hat{\overline{U}}_g(t)=e^{-i\omega{t}(n+\frac{3}{2})}\left\lbrace\cos(\overline{R}_{gn}t)\hat{\overline{I}}_g-i\sin(\overline{R}_{gn}t)\hat{\overline{\varepsilon}}_g\right\rbrace
\label{eq:evolop}
\end{equation}
which we substitute into equation Eq.~\eqref{eq:schtevol} and use the qubit state transition operations in Eq.~\eqref{eq:algop11} to obtain the time-evolving AJC qubit state vector in the form 
\begin{equation}
|\overline{\Psi}_{gn}(t)\rangle=e^{-i\omega{t}(n+\frac{3}{2})}\Big\{\cos(\overline{R}_{gn}t)|\psi_{gn}\rangle - i\sin(\overline{R}_{gn}t)|\overline{\phi}_{gn}\rangle \Big\}
\label{eq:antievolve}
\end{equation}
This time evolving state vector  describes Rabi oscillations between the basic qubit states $|\psi_{gn}\rangle$ and $|\overline{\phi}_{gn}\rangle$ at Rabi frequency $\overline{R}_{gn}$.

In order to determine the length of the Bloch vector associated with the state vector in Eq.~\eqref{eq:antievolve}, we introduce the density 
operator
\begin{subequations}
\begin{equation}
\hat{\overline{\rho}}_{gn}(t)=|\overline{\Psi}_{gn}(t)\rangle\langle{\overline{\Psi}_{gn}}(t)|\label{eq:den1}
\end{equation}
which we expand to obtain
\begin{eqnarray}
\hat{\overline{\rho}}_{gn}(t) &=& 
\cos^2(\overline{R}_{gn}t)|\psi_{gn}\rangle\langle\psi_{gn}|+\frac{i}{2}\sin(2\overline{R}_{gn}t)|\psi_{gn}\rangle\langle\overline{\phi}_{gn}|
\nonumber \\
&&
-\frac{i}{2}\sin(2\overline{R}_{gn}t)|\overline{\phi}_{gn}\rangle\langle\psi_{gn}|+\sin^2(\overline{R}_{gn}t)|\overline{\phi}\rangle\langle\overline{\phi}| \;.
\nonumber \\ 
&&
\label{eq:blochvec}
\end{eqnarray}
Defining the coefficients of the projectors in Eq.~\eqref{eq:blochvec} as
\begin{eqnarray}
\overline{\rho}_{gn}^{11}(t)&=&\cos^2(\overline{R}_{gn}t)
\quad;\quad
\overline{\rho}_{gn}^{12}(t)=\frac{i}{2}\sin(2\overline{R}_{gn}t)
\nonumber \\ 
\overline{\rho}_{gn}^{21}(t)&=& 
-\frac{i}{2}\sin(2\overline{R}_{gn}t)
\quad;\quad
\overline{\rho}_{gn}^{22}(t)=\sin^2(\overline{R}_{gn}t)
\quad
\label{eq:elementsdop}
\end{eqnarray}
and interpreting the coefficients in Eq.~\eqref{eq:elementsdop} as elements of a 
$2\times2$ density matrix $\overline{\rho}_{gn}(t)$, 
which we express in terms of standard Pauli 
operator matrices $I$, $\sigma_x$, $\sigma_y$ 
and $\sigma_z$ as
\begin{equation}
\overline{\rho}_{gn}(t)=
\begin{pmatrix}
\overline{\rho}_{gn}^{11}(t) & \overline{\rho}_{gn}^{12}(t)\\
\overline{\rho}_{gn}^{21}(t) & \overline{\rho}_{gn}^{22}(t)\\
\end{pmatrix} 
=\frac{1}{2}\left(I+\vec{\overline{\rho}}_{gn}(t)\cdot\vec{\sigma}\right) 
\end{equation}
where 
$\vec{\sigma}=\left(\sigma_x, \sigma_y, \sigma_z\right)$ 
is the Pauli matrix vector and we have introduced 
the time-evolving Bloch vector 
$\vec{\overline{\rho}}_{gn}(t)$ 
obtained in the form
\begin{equation}
\vec{\overline{\rho}}_{gn}(t)
=\left(\overline{\rho}_{gn}^x(t), \overline{\rho}_{gn}^y(t), \overline{\rho}_{gn}^z(t)\right)
\label{eq:blochvec0}
\end{equation}
with components defined as
\begin{eqnarray}
\overline{\rho}_{gn}^x(t) &=& 
\overline{\rho}_{gn}^{12}(t)+\overline{\rho}_{gn}^{21}(t)=0
\nonumber \\
\overline{\rho}_{gn}^y(t) &=& 
i\left(\overline{\rho}_{gn}^{12}(t)-\overline{\rho}_{gn}^{21}(t)\right)=-\sin(2\overline{R}_{gn}t)
\nonumber \\
\overline{\rho}_{gn}^z(t)&=& 
\overline{\rho}_{gn}^{11}(t)-\overline{\rho}_{gn}^{22}(t)=\cos(2\overline{R}_{gn}t)
\end{eqnarray} 
The Bloch vector in Eq.~\eqref{eq:blochvec0} takes the explicit form
\begin{equation}
\vec{\overline{\rho}}_{gn}(t)=
\Big
(0, -\sin(2\overline{R}_{gn}t), \cos(2\overline{R}_{gn}t)\Big)
\label{eq:blochvec1}
\end{equation}
which has unit length obtained easily as
\begin{equation}
|\vec{\overline{\rho}}_{gn}(t)|=1
\end{equation}
\end{subequations}
The property that the Bloch vector $\vec{\overline{\rho}}_{gn}(t)$ is of unit length (the Bloch sphere has unit radius), clearly shows that the general time evolving state vector $|\overline{\Psi}_{gn}(t)\rangle$ in Eq.~\eqref{eq:antievolve} is a pure state. 

We now proceed to demonstrate the time evolution of the Bloch vector $\vec{\overline{\rho}}_{gn}(t)$ which in effect describes the geometric configuration of states. We have adopted class 4 Bloch-sphere entanglement of a quantum rank-2 bipartite state \cite{boyer2017geometry, regula2016entanglement} to bring a clear visualization of this interaction. In this respect, we consider the specific example (which also applies to the general \textit{n}-photon case) of an atom initially in ground state $|g\rangle$ entering a cavity with the field mode starting off in an initial vacuum state $|0\rangle$, such that the initial atom-field state is $|g,0\rangle$. It is important to note that in the AJC interaction process the initial atom-field ground state $|g,0\rangle$ is an absolute ground state with both atom and field mode in the ground state $|g\rangle$, $|0\rangle$, in contrast to the commonly applied initial atom-field ground state $|e,0\rangle$ in the JC model where only the field mode $|0\rangle$ is in the ground state and the atom in the excited state $|e\rangle$.

In the specific example starting with an atom in the 
ground state $|g\rangle$ and the field mode in the 
vacuum state $|0\rangle$ the basic qubit state 
vectors $|\psi_{g0}\rangle$ and 
$|\overline{\phi}_{g0}\rangle$, together with the 
corresponding entanglement parameters, are obtained 
by setting $n=0$ in Eqs.~\eqref{eq:entsuptate} 
and \eqref{eq:parameters} in the form
\begin{eqnarray}
|\psi_{g0}\rangle &=& |g,0\rangle
\quad;\quad
|\overline{\phi}_{g0}\rangle = 
-\overline{c}_{g0}|g,0\rangle+\overline{s}_{g0}|e,1\rangle
\quad;
\nonumber \\ 
\overline{c}_{g0} &=& \frac{\overline{\delta}}{2\overline{R}_{g0}}\quad;\quad
\overline{s}_{g0} = \frac{2\lambda}{\overline{R}_{g0}}\quad;\quad
\overline{R}_{g0} =\frac{1}{2}\sqrt{16\lambda^2+\overline{\delta}^2}
\nonumber \\
|g,0\rangle &=& |g\rangle\otimes|0\rangle
\quad;\quad
|e,1\rangle=|e\rangle\otimes|1\rangle
\label{eq:initialstateeqs}
\end{eqnarray}
The corresponding Hamiltonian in Eq.~\eqref{eq:antijch2}  becomes ($n=0$)
\begin{equation}
\hat{\overline{H}}_g=\frac{3}{2}\hbar\omega\hat{\overline{I}}_g+\hbar\overline{R}_{g0}\hat{\overline{\varepsilon}}_g
\end{equation}
The time-evolving state vector in Eq.~\eqref{eq:antievolve} takes the form ($n=0$) 
\begin{equation}
|\overline{\Psi}_{g0}(t)\rangle = 
e^{-i\frac{3}{2}\omega{t}}\left\lbrace\cos(\overline{R}_{g0}t)|\psi_{g0}\rangle-i\sin(\overline{R}_{g0}t)|\overline{\phi}_{g0}\rangle\right\rbrace
\label{eq:antievolv}
\end{equation}
which describes Rabi oscillations at frequency 
$\overline{R}_{g0}$ between the initial separable 
qubit state vector $|\psi_{g0}\rangle$ and the 
entangled qubit state vector 
$|\overline{\phi}_{g0}\rangle$.

The Rabi oscillation process is best 
described by the corresponding Bloch vector which 
follows from Eq.~\eqref{eq:blochvec1} in the form ($n=0$) 
\begin{equation}
\vec{\overline{\rho}}_{g0}(t)=\left(0, -\sin(2\overline{R}_{g0}t), \cos(2\overline{R}_{g0}t)\right)
\label{eq:ajctevolvebloch}
\end{equation}
The time evolution of this Bloch vector reveals that the Rabi oscillations between the basic qubit state vectors 
$|\psi_{g0}\rangle$, $|\overline{\phi}_{g0}\rangle$ 
describe circles on which the states are distributed 
on the Bloch sphere as we demonstrate in 
Fig.~\ref{fig:AJC} below.
 
\begin{figure}
\centering
\includegraphics[scale=0.75]{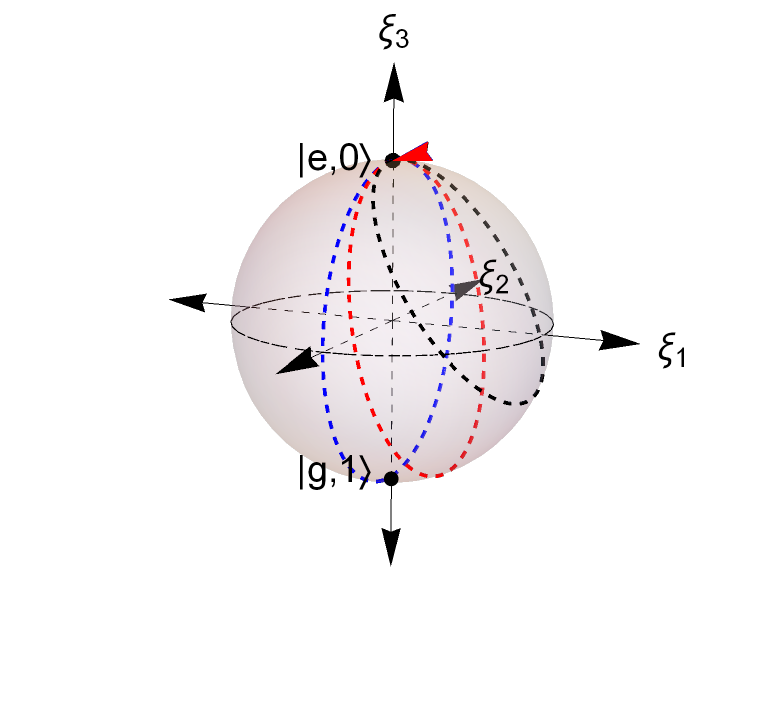} 
\vspace*{-2.50cm}
\caption{Rabi oscillations in JC interaction mechanism. 
Here, blue circle is at resonance with 
detuning $\delta=\omega_0-\omega=0$, red circle 
is for detuning $\delta=\lambda$ and black circle for detuning $\delta=3\lambda$.
}
\label{fig:JC}
\end{figure}

In Fig.~\ref{fig:AJC} we have plotted the AJC Rabi oscillation process with respective Rabi frequencies $\overline{R}_{g0}$ determined according to Eq.~\eqref{eq:initialstateeqs} for various values of sum frequency $\overline{\delta}=\omega_0+\omega$. We have provided a comparison 
with plots of the corresponding JC process 
in Fig.~\ref{fig:JC}. 

To facilitate the desired comparison of the AJC Rabi oscillation process with the standard JC Rabi oscillation process plotted in Fig.~\ref{fig:JC}, we substitute the redefinition $\overline{\delta}=\omega_0+\omega=\delta+2\omega$ to express the Rabi frequency $\overline{R}_{g0}$ in Eq.~\eqref{eq:initialstateeqs} in the form
\begin{equation}
\overline{R}_{g0}=\frac{1}{2}\sqrt{16\lambda^2+(\delta+2\omega)^2} \,.
\label{eq:newrabi}
\end{equation}

In the present work, we have chosen the field mode 
frequency $\omega=2\lambda$ ($\lambda=0.5\omega$) such that for both AJC 
and JC processes we vary only the detuning frequency 
$\delta=\omega_0-\omega$. The resonance case 
$\delta=0$ in the JC interaction now means $\overline{\delta}=2\omega=4\lambda$ in the 
AJC interaction. 

For various values of $\delta=\lambda\,,\,3\lambda\,,\,0$, we 
use the general time evolving state vector in Eq.~\eqref{eq:antievolv}, with $\overline{R}_{g0}$ as defined in Eq.~\eqref{eq:newrabi} to determine the coupled qubit state vectors $|\psi_{g0}\rangle$\,,\, $|\overline{\phi}_{g0}\rangle$  in Eq.~\eqref{eq:initialstateeqs} by setting $\overline{R}_{g0}t=\frac{\pi}{2}$, describing half cycle of Rabi oscillation as presented below. In each case we have an accumulated global phase factor which does not affect measurement results \cite{nielsen2011quantum, rieffel2011quantum, van2014quantum}, but we have maintained them here in Eqs.~\eqref{eq:transition1}\,-\,\eqref{eq:transition6} to explain the continuous time evolution over one cycle.

\begin{subequations}
\begin{equation*}
\delta=\lambda\quad;\quad\overline{\delta}=5\lambda\,:
\end{equation*}

\begin{equation}
|g,0\rangle \rightarrow  
 {e}^{-i\pi\frac{79}{82}}\left\lbrace-\frac{5}{\sqrt{41}}|g,0\rangle+\frac{4}{\sqrt{41}}|e,1\rangle\right\rbrace\rightarrow 
 {e}^{-i\pi\frac{79}{41}}|g,0\rangle 
\label{eq:transition1}
 \end{equation}

\begin{equation*}
\delta=3\lambda\quad;\quad\overline{\delta}=7\lambda\,:
\end{equation*}

\begin{equation}
|g,0\rangle \rightarrow 
{e}^{-i\pi\frac{113}{130}}\left\lbrace-\frac{7}{\sqrt{65}}|g,0\rangle+\frac{4}{\sqrt{65}}|e,1\rangle\right\rbrace\rightarrow{e}^{-i\pi\frac{113}{65}}|g,0\rangle
\label{eq:transition3}
\end{equation}

\begin{equation*}
\delta=0\quad;\quad\overline{\delta}=4\lambda\,:
\end{equation*}

\begin{equation}
|g,0\rangle\rightarrow{e}^{-i\pi}\left\lbrace{-\frac{1}{\sqrt{2}}}|g,0\rangle+\frac{1}{\sqrt{2}}|e,1\rangle\right\rbrace\rightarrow {e}^{-i\pi{2}}|g,0\rangle
\label{eq:transition6}
\end{equation}
\end{subequations}

The AJC Rabi oscillations for cases $\delta=\lambda\,,3\lambda\,,0$ are plotted as red, black and blue circles in Fig.~\ref{fig:AJC}, while the corresponding plots in the JC process are provided in Fig.~\ref{fig:JC} as a comparison. Here, Fig.~\ref{fig:AJC} is a Bloch sphere
entanglement \cite{boyer2017geometry} that corresponds to a 2-dimensional subspace of $\mathbb{C}^2 \hspace{0.5mm}\otimes\hspace{0.5mm}\mathbb{C}^2$  Span$\left\lbrace|g,0\rangle \hspace{0.5mm} , \hspace{0.5mm} -\overline{c}_{g0}|g,0\rangle+\overline{s}_{g0}|e,1\rangle\right\rbrace$ with $\overline{c}_{g0}=\frac{\overline{\delta}}{2\overline{R}_{g0}}$ and $\overline{s}_{g0}=\frac{2\lambda}{\overline{R}_{g0}}$ while Fig.~\ref{fig:JC} is a Bloch sphere entanglement corresponding to a 2-dimensional subspace of $\mathbb{C}^2\hspace{0.5mm}\otimes\hspace{0.5mm}\mathbb{C}^2$ Span$\left\lbrace|e,0\rangle \hspace{0.5mm}, \hspace{0.5mm}c_{e0}|e,0\rangle+s_{e0}|g,1\rangle\right\rbrace$ with $c_{e0}=\frac{\delta}{2R_{e0}}$ and $s_{e0}=\frac{2\lambda}{R_{e0}}$, where we recall that, in the JC interaction the initial atom-field ground state with the field mode in the vacuum state is $|e,0\rangle$.
\begin{figure}
\centering
\includegraphics[scale=0.75]{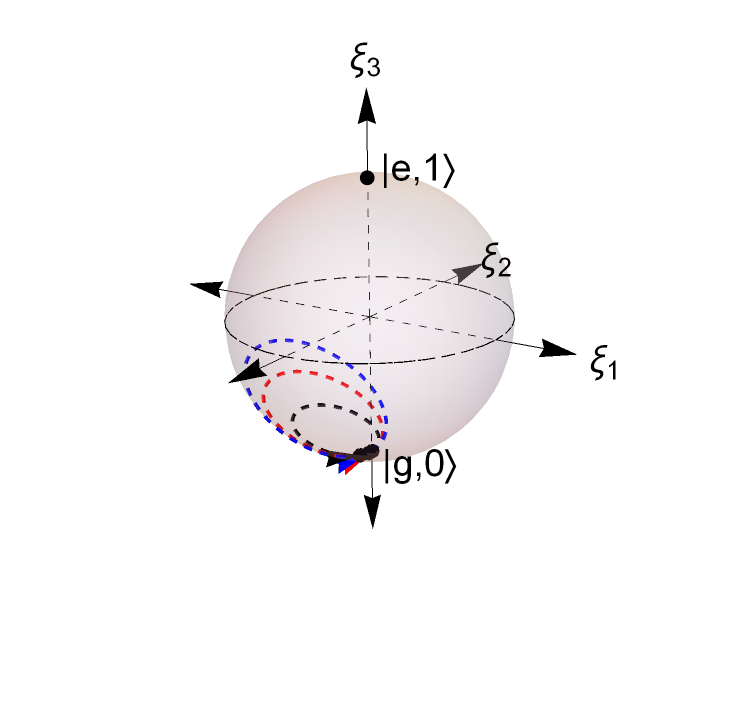}
\vspace*{-2.50cm} 
\caption{
Rabi oscillations in AJC interaction mechanism. 
The Rabi oscillations for values of sum 
frequencies are shown by red ($\overline{\delta}=5\lambda\,;\,\delta=\lambda$), black ($\overline{\delta}=7\lambda\,;\; \delta=3\lambda$) and blue 
($\overline{\delta}=4\lambda \,;\, \delta=\omega_0-\omega=0$).}
\label{fig:AJC}
\end{figure}

\begin{figure}
\centering
\includegraphics[scale=0.75]{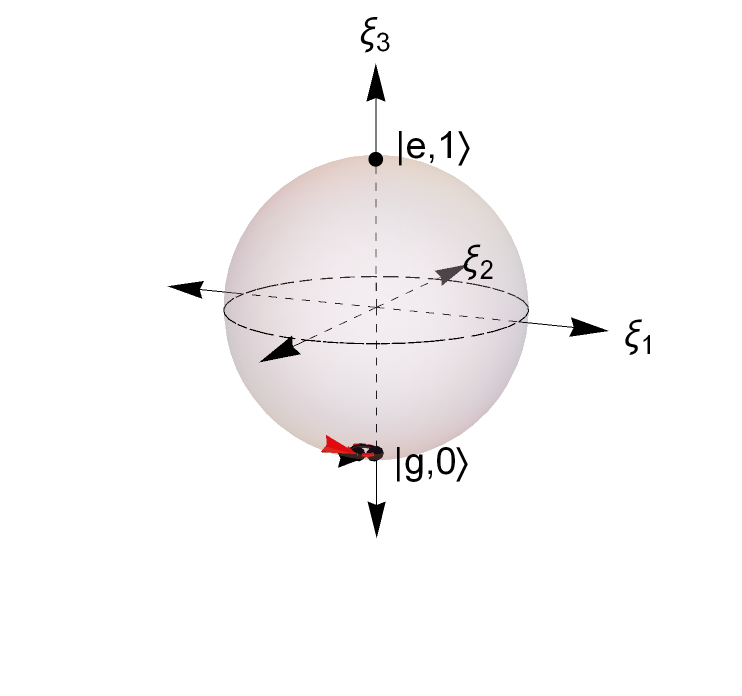} 
\vspace*{-2.50cm}
\caption{Rabi oscillations in AJC interaction mechanism. 
The Rabi oscillations for values of sum 
frequencies are shown by red ($\overline{\delta}=21\lambda\hspace{1mm};\hspace{1mm}\delta=\lambda$) and black ($\overline{\delta}=23\lambda\hspace{1mm};\hspace{1mm}\delta=3\lambda$).} 
\label{fig:AJCHIGH}
\end{figure}

In Fig.~\ref{fig:AJC} we observe:
\begin{enumerate}
\item[(i)] that due to the larger sum frequency $\overline{\delta}=\delta+2\omega$ in the AJC interaction process as compared to the detuning frequency $\delta$ in the JC interaction process, the Rabi oscillation circles in the much faster AJC process are much smaller compared to the corresponding Rabi oscillation circles in the slower JC interaction process. This effect is in agreement with the assumption usually adopted to drop the AJC interaction components in the rotating wave approximation (RWA), noting that the fast oscillating AJC process averages out over time. We have demonstrated the physical property that the size of the Rabi oscillations curves decreases with increasing Rabi oscillation frequency by plotting the AJC oscillation curves for a considerably larger Rabi frequency $\overline{R}_{g0}$  where we have set the field mode frequency $\omega=10\lambda (\lambda=0.1\omega)$ in Fig.~\ref{fig:AJCHIGH}. It is clear in Fig.~\ref{fig:AJCHIGH} that for this higher value of the Rabi frequency $\overline{R}_{g0}$  the Rabi oscillation curves almost converge to a point-like form;

\item[(ii)] that Rabi oscillations in the AJC interaction process as demonstrated in Fig.~\ref{fig:AJC} occur in the left hemisphere of the Bloch sphere while in the JC interaction process the oscillations occur in the right hemisphere as demonstrated in Fig.~\ref{fig:JC}. This demonstrates an important physical property that the AJC interaction process occurs in the reverse sense relative to the JC interaction process;

\item[(iii)] an interesting feature that appears at resonance specified by $\delta=0$. While in the JC model plotted in Fig~\ref{fig:JC} the Rabi oscillation at resonance $\delta=0$ (blue circle) lies precisely on the \textit{yz}-plane normal to the equatorial plane,  the corresponding AJC Rabi oscillation (blue circle in Fig.~\ref{fig:AJC}) is at an axis away from the \textit{yz}-plane about the south pole of the Bloch sphere. This feature is due to the fact that the frequency detuning $\overline{\delta}=2\omega$ takes a non-zero value under resonance $\delta=0$ such that the AJC oscillations maintain their original forms even under resonance.
\end{enumerate} 

We note that the qubit state transitions described by the Bloch vector in the AJC process (Fig.~\ref{fig:AJC}) are blue-side band transitions characterized by the sum frequency $\overline{\delta}=\omega_0+\omega=\delta+2\omega$ according to the definition of the Rabi frequency $\overline{R}_{g0}$ in eq.~\eqref{eq:newrabi}.

The geometric configuration of the state space demonstrated on the Bloch-sphere in Fig.~\ref{fig:JC} determined using the approach in  \cite{omolo2019photospins} agrees precisely with that determined using the semi-classical approach in  \cite{enriquez2014note} corresponding to a 2-dimensional subspace of $\mathbb{C}^2$ Span $\left\lbrace|e\rangle\,,|g\rangle\right\rbrace$. In the approach \cite{enriquez2014note}, at resonance where detuning $\delta=0$ the atomic population is inverted from $|e\rangle$ to $|g\rangle$ and the Bloch-vector $\vec{r}=(\sin(\theta)\cos(\phi)\,,\sin(\theta) \sin(\phi)\,,\cos(\theta))$ describes a path along the \textit{yz}-plane on the Bloch-sphere. For other values of detuning, the atom evolves  from $|e\rangle$ to a linear superposition of $|e\rangle$ and $|g\rangle$ and back to $|e\rangle$ and the Bloch-vector $\vec{r}$ describes a circle about the north pole of the Bloch-sphere.
%
%
\section{Entanglement properties}
\label{sec:entquant}
In quantum information, it is of interest to measure or quantify the entanglement of states. In this paper we apply the von Neumann entropy as a measure of entanglement. The von Neumann entropy  \cite{von2018mathematical, wootters2001entanglement, abdel2011dynamics, bennett1996mixed, bennett1996concentrating} of a quantum state $\hat{\rho}$ is defined as

\begin{equation}
S(\hat{\rho})=-tr\left(\hat{\rho}\log\hat{\rho}\right)=-\sum_i\lambda_i\log\lambda_i
\label{eq:von}
\end{equation}
where the logarithm is taken to base \textit{d}, \textit{d} being the dimension of the Hilbert space containing $\hat{\rho}$ and $\lambda_is$ are the eigenvalues that diagonal $\hat{\rho}$. It follows that $0\leqslant{S(\hat{\rho})}\leqslant{1}$, where $S(\hat{\rho})=0$ if and only if $\hat{\rho}$ is a pure state.

Further, the von Neumann entropy of the reduced density matrices of a bipartite pure state  
$\hat{\rho}_{AB}=|\psi_{AB}\rangle\langle\psi_{AB}|$ is 
a good and convenient entanglement measure 
E($\hat{\rho}_{AB}$). The entanglement measure defined 
as the entropy of either of the quantum subsystem  is obtained as
\begin{equation}
E(\hat{\rho}_{AB})=-tr(\hat{\rho}_A\log_2\hat{\rho}_A)=-tr(\hat{\rho}_B\log_2\hat{\rho}_B)
\label{eq:thevon}
\end{equation}
where for all states we have $0\leq{E(\hat{\rho}_{AB})}\leq{1}$. Here the limit $0$ is achieved if the pure state is a product $|\psi\rangle=|\psi_A\rangle\otimes|\psi_B\rangle$ and $1$ is achieved for maximally entangled states, noting that the reduced density matrices are maximally mixed states. 

In this section we analyse the entanglement properties of the qubit state vectors and the dynamical evolution of entanglement generated in the AJC interaction.
%
%
\subsection{Entanglement analysis of basic qubit state vectors $|\psi_{g0}\rangle$ and $|\overline{\phi}_{g0}\rangle$}
\label{sec:entanalysis}
Let us start by considering the entanglement properties of the initial state $|\psi_{g0}\rangle$ which according to the definition in Eq.~\eqref{eq:initialstateeqs} is a separable pure state. The density operator of the qubit state vector $|\psi_{g0}\rangle=|g,0\rangle$ is obtained as
\begin{subequations}
\begin{equation}
\hat{\rho}_{g0}=|g,0\rangle\langle{g,0}|
\label{eq:binitdens}
\end{equation}
Using the definition $|g,0\rangle=|g\rangle\otimes|0\rangle$, we take the partial trace of 
$\hat{\rho}_{g0}$ in Eq.~\eqref{eq:binitdens} with 
respect to the field mode and atom states respectively, 
to obtain the respective atom and field reduced 
density operators $\hat{\rho}_A$, $\hat{\rho}_F$ in 
the form (subscripts $A\equiv{atom}$ and $F\equiv{field}$)
\begin{equation}
\hat{\rho}_A=tr_F(\hat{\rho}_{g0})=|g\rangle\langle{g}| \quad;\quad\hat{\rho}_F=tr_A(\hat{\rho}_{g0})=|0\rangle\langle{0}|
\label{eq:redbinitdens}
\end{equation}
which take explicit $2\times2$ matrix forms
\begin{equation}
\hat{\rho}_A=
\begin{pmatrix}
0 & 0\\
0 & 1\\
\end{pmatrix}\quad;\quad
\hat{\rho}_F=
\begin{pmatrix}
1 & 0\\
0 & 0\\
\end{pmatrix}
\label{eq:puremat}
\end{equation}
The trace of $\hat{\rho}_A$,\,$\hat{\rho}_A^2$ and $\hat{\rho}_F$,\,$\hat{\rho}_F^2$ of the matrices in Eq.~\eqref{eq:puremat} are
\begin{equation}
tr(\hat{\rho}_A)=tr(\hat{\rho}_A^2)=1\quad;\quad{tr}(\hat{\rho}_F)=tr(\hat{\rho}_F^2)=1
\label{eq:ptrace}
\end{equation}
The unit trace determined in Eq.~\eqref{eq:ptrace} proves that the initial qubit state vector $|\psi_{g0}\rangle=|g,0\rangle$ is a pure state.

Next, we substitute the matrix form of $\hat{\rho}_A$ and $\hat{\rho}_F$ from Eq.~\eqref{eq:puremat} into Eq.~\eqref{eq:thevon} to obtain equal von Neumann entanglement entropies
\begin{equation}
E(\hat{\rho}_{g0})=S(\hat{\rho}_A)=S(\hat{\rho}_F)=0
\end{equation}
\end{subequations}
which together with the property in Eq.~\eqref{eq:ptrace} quantifies the initial qubit state vector $|\psi_{g0}\rangle=|g,0\rangle$ as a pure separable state, agreeing with the definition in Eq.~\eqref{eq:initialstateeqs}.
   
We proceed to determine the entanglement properties of the (transition) qubit state vector $|\overline{\phi}_{g0}\rangle$ defined in Eq.~\eqref{eq:initialstateeqs}. For parameter values $\delta=\lambda\hspace{1mm},\hspace{1mm}\overline{\delta}=5\lambda$ we ignore the phase factor in Eq.~\eqref{eq:transition1}, to write the transition qubit state vector in the form 
\begin{subequations}
\begin{equation}
\delta=\lambda\quad;\quad\overline{\delta}=5\lambda\,:\quad|\overline{\phi}_{g0}\rangle=-\frac{5}{\sqrt{41}}|g,0\rangle+\frac{4}{\sqrt{41}}|e,1\rangle
\label{eq:antpolstate}
\end{equation}
The corresponding density operator of the state in Eq.~\eqref{eq:antpolstate} is
\begin{eqnarray}
\hat{\overline{\rho}}_{g0} &=& \frac{25}{41}|g,0\rangle\langle{g,0}|-\frac{20}{41}|g,0\rangle\langle{e,1}|-\frac{20}{41}|e,1\rangle\langle{g,0}| 
\nonumber \\
&& +\frac{16}{41}|e,1\rangle\langle{e,1}|
\label{eq:densantpolstate}
\end{eqnarray}
which takes the explicit $4\times4$ matrix form
\begin{equation}
\hat{\overline{\rho}}_{g0}=
\begin{pmatrix}
0 & 0 & 0 & 0\\
0 & \frac{16}{41} & -\frac{20}{41} & 0\\
0 & -\frac{20}{41} & \frac{25}{41} & 0\\
0 & 0 & 0 & 0\\
\end{pmatrix}
\end{equation}
with eigenvalues $\lambda_1=1$,\, $\lambda_2=0$,\, $\lambda_3=0$,\, $\lambda_4=0$. Applying  Eq.~\eqref{eq:von}, its von Neumann entropy 
\begin{equation}
S(\hat{\overline{\rho}}_{g0})=0
\label{eq:entzero1}
\end{equation}
quantifying the state $|\overline{\phi}_{g0}\rangle$  in Eq.~\eqref{eq:antpolstate} as a bipartite pure state.

Taking the partial trace of $\hat{\overline{\rho}}_{g0}$ in Eq.~\eqref{eq:densantpolstate} with respect to the field mode and atom states respectively, we obtain the respective atom and field reduced density operators $\hat{\overline{\rho}}_A$,\,$\hat{\overline{\rho}}_F$ together with their squares in the form 
\begin{eqnarray}
\hat{\overline{\rho}}_A &=& tr_F(\hat{\overline{\rho}}_{g0})=\frac{25}{41}|g\rangle\langle{g}|+\frac{16}{41}|e\rangle\langle{e}|\quad;\quad
\nonumber \\
\hat{\overline{\rho}}_A^2 &=& \frac{625}{1681}|g\rangle\langle{g}|+\frac{256}{1681}|e\rangle\langle{e}|
\nonumber \\ 
\hat{\overline{\rho}}_F &=& tr_A(\hat{\overline{\rho}}_{g0})=\frac{25}{41}|0\rangle\langle{0}|+\frac{16}{41}|1\rangle\langle{1}|\quad;\quad
\nonumber \\
\hat{\overline{\rho}}_F^2 &=& \frac{625}{1681}|0\rangle\langle{0}|+\frac{256}{1681}|1\rangle\langle{1}|
\label{eq:redantpolstae}
\end{eqnarray}
The trace of $\hat{\overline{\rho}}_A^2$ and $\hat{\overline{\rho}}_F^2$ in Eq.~\eqref{eq:redantpolstae} gives
\begin{equation}
tr(\hat{\overline{\rho}}_A^2)=tr(\hat{\overline{\rho}}_F^2)=\frac{881}{1681}<1
\label{eq:trc3}
\end{equation}
demonstrating that $\hat{\overline{\rho}}_A$ and $\hat{\overline{\rho}}_F$ are mixed states. To quantify the mixedness we determine the length of the Bloch vector along the $z$-axis as follows
\begin{equation}
r_z=tr(\hat{\overline{\rho}}_{A}\hat{\sigma}_z)=tr(\hat{\overline{\rho}}_{F}\hat{\sigma}_z)=\frac{9}{41}
\label{eq:lblch2}
\end{equation}
which shows that the reduced density operators $\hat{\overline{\rho}}_A$,\,$\hat{\overline{\rho}}_F$ 
are non-maximally mixed states.

The eigenvalues $\left(\lambda_1,\,\lambda_2\right)$ of $\hat{\overline{\rho}}_A$ and $\hat{\overline{\rho}}_F$ are $\left(\frac{16}{41},\,\frac{25}{41}\right)$ and $\left(\frac{25}{41},\,\frac{16}{41}\right)$ respectively, which on substituting into Eq.~\eqref{eq:von}, gives 
equal von Neumann entanglement entropies
\begin{eqnarray}
E(\hat{\overline{\rho}}_{g0}) &=& S(\hat{\overline{\rho}}_A)=S(\hat{\overline{\rho}}_F)
\nonumber \\
&=& -\frac{16}{41}\log_2\left(\frac{16}{41}\right)-\frac{25}{41}\log_2\left(\frac{25}{41}\right)=0.964957
\nonumber \\
\label{eq:entropy9}
\end{eqnarray}
\end{subequations}
Taking the properties in  Eqs.~\eqref{eq:entzero1}, \eqref{eq:trc3}\,-\,\eqref{eq:entropy9} together 
clearly characterizes the qubit state 
$|\overline{\phi}_{g0}\rangle$ in 
Eq.~\eqref{eq:antpolstate} as an entangled bipartite 
pure state. However, since 
$S(\hat{\overline{\rho}}_A)=S(\hat{\overline{\rho}}_F)<1$ 
the state is not maximally entangled. Similarly, the transition qubit state vector 
$|\overline{\phi}_{g0}\rangle=-\frac{7}{\sqrt{65}}|g,0\rangle+\frac{4}{\sqrt{65}}|e,1\rangle$ 
obtained for 
$\delta=3\lambda,\,\overline{\delta}=7\lambda$ in 
Eq.~\eqref{eq:transition3} is an entangled bipartite 
pure state, but not maximally entangled.

Finally, we consider the resonance case $\delta=0$, characterized by $\overline{\delta}=4\lambda$ in the AJC. Ignoring the phase factor in Eq.~\eqref{eq:transition6} 
the transition qubit state vector 
$|\overline{\phi}_{g0}\rangle$ takes the form
\begin{subequations}
\begin{equation}
\delta=0\quad;\quad\overline{\delta}=4\lambda\,:\quad|\overline{\phi}_{g0}\rangle=-\frac{1}{\sqrt{2}}|g,0\rangle+\frac{1}{\sqrt{2}}|e,1\rangle
\label{eq:entatresonance}
\end{equation}
The corresponding density operator of the state in Eq.~\eqref{eq:entatresonance} is
\begin{equation}
\hat{\overline{\rho}}_{g0}=\frac{1}{2}|g,0\rangle\langle{g,0}|-\frac{1}{2}|g,0\rangle\langle{e,1}|-\frac{1}{2}|e,1\rangle\langle{g,0}|+\frac{1}{2}|e,1\rangle\langle{e,1}|
\label{eq:densentagresonance}
\end{equation}
which takes the explicit $4\times{4}$ matrix form
\begin{equation}
\hat{\overline{\rho}}_{g0}=
\begin{pmatrix}
0 & 0 & 0 & 0\\
0 & \frac{1}{2} & -\frac{1}{2} & 0\\
0 & -\frac{1}{2} & \frac{1}{2} & 0\\
0 & 0 & 0 & 0\\
\end{pmatrix}
\end{equation}
with eigenvalues $\lambda_1=1\,,\, 
\lambda_2=0\,,\,\lambda_3=0\,,\,\lambda_4=0$. Applying eq.~\eqref{eq:von} its von Neumann entropy
\begin{equation}
S(\hat{\overline{\rho}}_{g0})=0
\label{eq:zeroentropy}
\end{equation}
quantifying the state in Eq.~\eqref{eq:entatresonance} as a bipartite pure state. 

Taking the partial trace of $\hat{\overline{\rho}}_{g0}$ in Eq.~\eqref{eq:densentagresonance} with respect to the field mode and atom states respectively, we obtain the respective atom and field reduced density operators $\hat{\overline{\rho}}_A$,\,$\hat{\overline{\rho}}_F$ together with their squares in the form 
\begin{eqnarray}
\hat{\overline{\rho}}_A &=& tr_F(\hat{\overline{\rho}}_{g0})=\frac{1}{2}|g\rangle\langle{g}|+\frac{1}{2}|e\rangle\langle{e}|\quad;\quad
\nonumber \\
\hat{\overline{\rho}}_A^2 &=& \frac{1}{4}|g\rangle\langle{g}|+\frac{1}{4}|e\rangle\langle{e}|
\nonumber \\
\hat{\overline{\rho}}_F &=& tr_A(\hat{\overline{\rho}}_{g0})=\frac{1}{2}|0\rangle\langle{0}|+\frac{1}{2}|1\rangle\langle{1}|\quad;\quad
\nonumber \\
\hat{\overline{\rho}}_F^2 &=& \frac{1}{4}|0\rangle\langle{0}|+\frac{1}{4}|1\rangle\langle{1}|
\label{eq:redantpolstae33}
\end{eqnarray}
The trace of $\hat{\overline{\rho}}_A^2$ and $\hat{\overline{\rho}}_F^2$ in 
Eq.~\eqref{eq:redantpolstae33} is
\begin{equation}
tr(\hat{\overline{\rho}}_A^2)=tr(\hat{\overline{\rho}}_F^2)=\frac{1}{2}<1
\label{eq:dsquare}
\end{equation} 
which reveals that the reduced density operators  $\hat{\overline{\rho}}_{A}$, $\hat{\overline{\rho}}_F$ are mixed states. To quantify the mixedness, we determine the length of the Bloch vector along the \textit{z}-axis as follows
\begin{equation}
r_z=tr(\hat{\overline{\rho}}_A\hat{\sigma}_z)=tr(\hat{\overline{\rho}}_F\hat{\sigma}_z)=0
\label{eq:lblch}
\end{equation} 
showing that the reduced density operators $\hat{\overline{\rho}}_A$ and $\hat{\overline{\rho}}_F$ are maximally mixed states.

The eigenvalues $(\lambda_1, \lambda_2)$ of $\hat{\overline{\rho}}_A$ and $\hat{\overline{\rho}}_F$ are $(\frac{1}{2}, \frac{1}{2})$ respectively which on substituting into Eq.~\eqref{eq:von}, gives equal von Neumann entanglement entropies
\begin{eqnarray}
E(\hat{\overline{\rho}}_{g0}) &=& S(\hat{\overline{\rho}}_A)=S(\hat{\overline{\rho}}_F)
\nonumber \\
&=&-\frac{1}{2}\log_2\left(\frac{1}{2}\right)-\frac{1}{2}\log_2\left(\frac{1}{2}\right)=1
\label{eq:thisunitent}
\end{eqnarray}
\end{subequations}
The unit entropy determined in Eq.~\eqref{eq:thisunitent} together with the properties in 
Eqs.~\eqref{eq:zeroentropy}\,-\,\eqref{eq:lblch} quantifies the transition qubit state determined at resonance $\delta=0$ in Eq.~\eqref{eq:entatresonance} (or Eq.~\eqref{eq:transition6}) as a maximally entangled bipartite pure state. Due to this maximal entanglement property, we shall use the resonance transition qubit 
state~$|\overline{\phi}_{g0}\rangle$ in Eq.~\eqref{eq:entatresonance} to implement teleportation by entanglement swapping protocol in Sec.~\ref{sec:teleportation} below. 

Similar proof of entanglement of the AJC qubit states is easily achieved for all possible values of sum frequency parameter $\overline{\delta}=\omega_0+\omega$, confirming that in the initial vacuum-field AJC interaction, reversible transitions occur only  between a pure initial separable qubit state vector $|\psi_{g0}\rangle$ and a pure entangled qubit state vector $|\overline{\phi}_{g0}\rangle$. This property of Rabi oscillations between an initial separable state and an entangled transition qubit state occurs in the general AJC interaction described by the general time evolving state vector $|\overline{\Psi}_{gn}(t)\rangle$ in Eq.~\eqref{eq:antievolve}.
%
%
\subsection{Entanglement evolution}
\label{sec:entangdeg}

Let us consider the general dynamics of AJC interaction described by the general time-evolving qubit state vector $|\overline{\Psi}_{gn}(t)\rangle$ in Eq.~\eqref{eq:antievolve}. Substituting $|\overline{\Psi}_{gn}(t)\rangle$ from Eq.~\eqref{eq:antievolve} into Eq.~\eqref{eq:den1} and using the definitions of $|\psi_{gn}\rangle$, $|\overline{\phi}_{gn}\rangle$ in Eq.~\eqref{eq:entsuptate} the density operator takes the form

\begin{widetext}
  \begin{flalign}
     \displaystyle 
\hat{\overline{\rho}}_{gn}(t)=\left\lbrace\cos^2(\overline{R}_{gn}t)+\overline{c}_{gn}^2\sin^2(\overline{R}_{gn}t)\right\rbrace|g,n\rangle\langle{g,n}|+\left\lbrace{i}~\overline{s}_{gn}\cos(\overline{R}_{gn}t)\sin(\overline{R}_{gn}t)-\overline{c}_{gn}\overline{s}_{gn}\sin^2(\overline{R}_{gn}t)\right\rbrace|g,n\rangle\langle{e,n+1}|
\nonumber \\
+\left\lbrace{-i}~\overline{s}_{gn}\cos(\overline{R}_{gn}t)\sin(\overline{R}_{gn}t)-\overline{c}_{gn}\overline{s}_{gn}\sin^2(\overline{R}_{gn}t)\right\rbrace|e,n+1\rangle\langle{g,n}|+\left\lbrace\overline{s}_{gn}^2\sin^2(\overline{R}_{gn}t)\right\rbrace|e,n+1\rangle\langle{e,n+1}|
\label{eq:densop}
   \end{flalign} 
\end{widetext} 
The reduced density operator of the atom is determined by tracing over the field states, thus taking the form
\begin{equation}
\hat{\overline{\rho}}_A(t)=P_g(t)|g\rangle\langle{g}|+P_e(t)|e\rangle\langle{e}|
\label{eq:redoperator1}
\end{equation}
after introducing the general time evolving atomic state probabilities $P_g(t)$, $P_e(t)$ obtained as
\begin{eqnarray}
P_g(t) &=& \cos^2(\overline{R}_{gn}t)+\overline{c}_{gn}^2\sin^2(\overline{R}_{gn}t)
\nonumber \\
P_e(t) &=& \overline{s}_{gn}^2\sin^2(\overline{R}_{gn}t)
\label{eq:probamps}
\end{eqnarray}
where the dimensionless interaction parameters $\overline{c}_{gn}$, $\overline{s}_{gn}$ are defined in Eq.~\eqref{eq:parameters} and the Rabi frequency takes the form
\begin{equation}
\overline{R}_{gn}=\frac{1}{2}\sqrt{16\lambda^2(n+1)+\overline{\delta}^2}
\label{eq:freshrabi}
\end{equation}
Expressing $\hat{\overline{\rho}}_A(t)$ in Eq.~\eqref{eq:redoperator1} in $2\times2$ matrix form

\begin{equation}
\hat{\overline{\rho}}_A(t)=
\begin{pmatrix}
P_e(t) & 0\\
0 & P_g(t)\\
\end{pmatrix}
\end{equation}
we determine the quantum system entanglement 
degree $E(t)$ defined in Eq.~\eqref{eq:thevon} as 
\begin{eqnarray}
E(t)&=&-tr(\hat{\overline{\rho}}_A(t)\log_2\hat{\overline{\rho}}_A(t)) 
\nonumber \\
&=& 
-tr
\left(
\begin{pmatrix}
P_e(t) & 0\\
0 & P_g(t)\\
\end{pmatrix}
\begin{pmatrix}
\log_2P_e(t) & 0\\
0 & \log_2P_g(t)\\
\end{pmatrix}
\right) 
\nonumber \\
\end{eqnarray}
which takes the final form
\begin{equation}
E(t)=-P_e(t)\log_2P_e(t)-P_g(t)\log_2P_g(t)
\label{eq:entropy}
\end{equation}
Using the definitions of the dimensionless 
parameters $\overline{c}_{gn}$, $\overline{s}_{gn}$ 
and the Rabi frequency $\overline{R}_{gn}$ in 
Eqs.~\eqref{eq:parameters}\,,\,\eqref{eq:freshrabi},  we evaluate the probabilities in Eq.~\eqref{eq:probamps} 
and plot the quantum system entanglement degree 
$E(\tau)$ in Eq.~\eqref{eq:entropy} against scaled 
time $\tau=\lambda{t}$ for arbitrarily chosen values 
of sum frequency 
$\overline{\delta}=2\lambda\,,\,6\lambda\,,\,8\lambda$ 
and photon number $n=1,\,2 ,\,3,\,6$ in 
Figs.~\ref{fig:graphA1}\,-\,\ref{fig:graphD} below.
\begin{figure}
\centering
\includegraphics[scale=0.6]{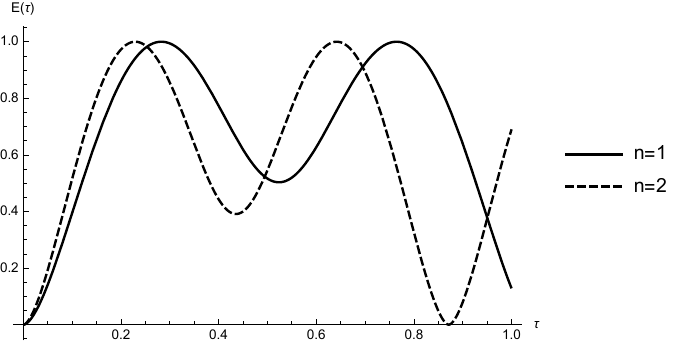}
\caption{\label{fig:graphA1}Degree of entanglement against scaled time for sum frequency $\overline{\delta}=2\lambda$ when $n=1$ and $n=2$}
\end{figure}
\begin{figure}
\centering
\includegraphics[scale=0.6]{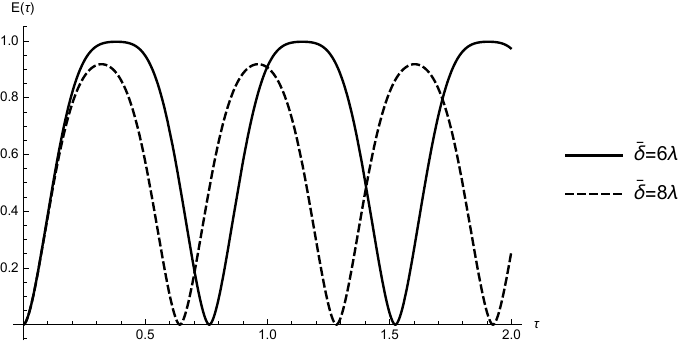}
\caption{\label{fig:graphB}Degree of entanglement against scaled time for sum frequency $\overline{\delta}=6\lambda$ and $\overline{\delta}=8\lambda$ when $n=1$}
\end{figure}
\begin{figure}
\centering
\includegraphics[scale=0.6]{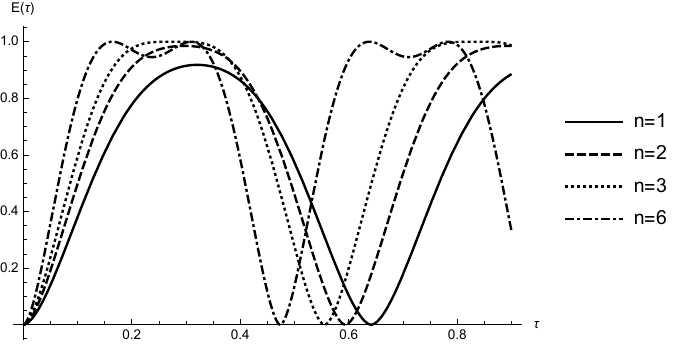}
\caption{\label{fig:graphD}Degree of entanglement against scaled time for sum frequency $\overline{\delta}=8\lambda$ when $n=1$, $n=2$, $n=3$ and $n=6$} 
\end{figure}

The graphs in Figs.~\ref{fig:graphA1}\,-\,\ref{fig:graphD}  show the effect of photon number $n$ and sum frequency $\overline{\delta}=\omega_0+\omega$ on the dynamical 
behavior of quantum entanglement measured by the von 
Neumann entropy $E(\tau)$ (min $E(\tau)=0$\,; max $E(\tau)=1$).  In the three figures, the phenomenon of entanglement sudden birth (ESB) and sudden death (ESD) is observed during the time evolution of entanglement similar to that observed in the JC model \cite{mohammadi2019time, liu2018nonlinear, zhang2018quantum}. In ESB there is an observed creation of entanglement where the initially un-entangled qubits are entangled after a very short time interval. For fairly low values of photon numbers $n$ and sum frequency $\overline{\delta}$ as demonstrated in Fig.~\ref{fig:graphA1} for $\overline{\delta}=2\lambda$ plotted when $n=1$, $n=2$, the degree of entanglement rises sharply to a maximum value of unity ($E(\tau)_{max}$) at an entangled state, stays at the maximum level for a reasonably short duration, decreases to a local minimum, then rises back to the maximum value before falling sharply to zero ($E(\tau)_{min}$) at the separable state. The local minimum disappears for larger values of sum frequency $\overline{\delta}\geq6\lambda$ at low photon number \textit{n} and re-emerge at high photon number $\textit{n}\geq4$ (see Fig.~\ref{fig:graphB} and Fig.~\ref{fig:graphD}) as examples.  However, in comparison to the resonance case $\delta=0$ in the JC model \cite{zhang2018quantum} we notice a long-lived entanglement at $E(\tau)_{max}=1$ in the cases of  $\overline{\delta}=6\lambda$ plotted when $n=1$ in Fig.~\ref{fig:graphB} and $\overline{\delta}=8\lambda$ plotted when $n=3$ in Fig.~\ref{fig:graphD}. The process of ESB and ESD then repeats periodically, consistent with Rabi oscillations between the qubit states.

In Fig.~\ref{fig:graphA1} and Fig.~\ref{fig:graphD} sum frequencies are kept constant at $\overline{\delta}=2\lambda$ and $\overline{\delta}=8\lambda$ respectively and photon number \textit{n} is varied in each case. We clearly see that the frequency of oscillation of $E(\tau)$ increases with an increase in photon number \textit{n}. This phenomenon in which the frequency of oscillation of $E(\tau)$ increases with an increase in photon number \textit{n} is also observed in the JC model  \cite{zhang2018quantum, liu2018nonlinear}. 

To visualize the effect of sum frequency parameter $\overline{\delta}$ on the dynamics of $E(\tau)$, we considered values of sum frequency set at $\overline{\delta}=6\lambda$ and $\overline{\delta}=8\lambda$ for photon number $n=1$ in Fig.~\ref{fig:graphB}. It is clear that the frequency of oscillation of $E(\tau)$ increases with an increase in sum frequency $\overline{\delta}=\omega_0+\omega$. In the JC model when detuning $\delta=\omega_0-\omega$ is set at off resonance $\delta\neq0$ results into a decrease in the frequency of oscillation of $E(\tau)$ as seen in \cite{liu2018nonlinear,zhang2018quantum, abdel2015effects} in comparison to the resonance case  $\delta=0$. 

Finally, for $\overline{\delta}=8\lambda$ plotted when $n=1$ in Fig.~\ref{fig:graphB} and in Fig.~\ref{fig:graphD} in comparison to $\overline{\delta}=6\lambda$ plotted when $n=1$ in Fig.~\ref{fig:graphB}, it is clear in Fig.~\ref{fig:graphB} that the degree of entanglement $E(\tau)$ decreases at a high value of sum frequency a phenomena similar to the JC model in \cite{abdel2015effects}. The observed decrease in degree of entanglement is due the property that the system loses its purity and the entropy decreases when the effect of sum frequency is considered for small number of photons \textit{n}. This is remedied when the effect of sum frequency is considered for higher photon numbers \textit{n} as shown in Fig.~\ref{fig:graphD}.

%
%
\section{Teleportation}
\label{sec:teleportation}

In the present work we consider an interesting case of quantum teleportation by applying entanglement swapping protocol (teleportation of entanglement) \cite{zukowski1993event, megidish2013entanglement, pan1998experimental, bose1998multiparticle}  where the teleported state is itself entangled. The state we want to teleport is a two-atom maximally entangled state in which we have assigned subscripts to distinguish the atomic qubit states in the form \cite{sousa2019selective}

\begin{equation}
|\varphi\rangle_{12}=\frac{1}{\sqrt{2}}(|e\rangle_1|g\rangle_2-|g\rangle_1|e\rangle_2)
\label{eq:twoatom}
\end{equation}
and it is in Alice's possession. In another location Bob is in possession of a maximally entangled qubit state $|\overline{\phi}_{g0}\rangle$ generated in the AJC interaction in Eq.~\eqref{eq:transition6} and expressed as

\begin{equation}
|\Phi\rangle_{3\textit{x}}=-\frac{1}{\sqrt{2}}|g\rangle_3|0\rangle_x+\frac{1}{\sqrt{2}}|e\rangle_3|1\rangle_x
\label{eq:transition6new}
\end{equation}
where we have also assigned subscripts to the qubits in Eq.~\eqref{eq:transition6new} to clearly distinguish them. 

An observer, Charlie, receives qubit-1 from Alice and 
qubit-$x$ from Bob. The entire state of the system 
\begin{subequations}
\begin{equation}
|\chi\rangle=|\varphi\rangle_{12}\otimes|\Phi\rangle_{3\textit{x}}
\end{equation}
which on substituting $|\varphi\rangle_{12}$ and $|\Phi\rangle_{3\textit{x}}$ from Eqs.~\eqref{eq:twoatom}, ~\eqref{eq:transition6new} and reorganizing takes the form

\begin{widetext}
\begin{flalign}
|\chi\rangle=\frac{1}{2}\Bigg[
|\Psi^+\rangle_{1\textit{x}}\left(\frac{|e\rangle_3|g\rangle_2 + 
|g\rangle_3|e\rangle_2}{\sqrt{2}}\right) + |\Psi^-\rangle_{1\textit{x}}\left(\frac{|e\rangle_3|g\rangle_2-|g\rangle_3|e\rangle_2}{\sqrt{2}}\right)
-|\Phi^-\rangle_{1\textit{x}}\left(\frac{|g\rangle_3|g\rangle_2-|e\rangle_3|e\rangle_2}{\sqrt{2}}\right) 
\nonumber \\
-|\Phi^+\rangle_{1\textit{x}}\left(\frac{|g\rangle_3|g\rangle_2+|e\rangle_3|e\rangle_2}{\sqrt{2}}\right)\Bigg]
\hfill
\label{eq:systems2}
   \end{flalign} 
\end{widetext} 
\end{subequations}
after introducing the emerging Bell states obtained as
\begin{eqnarray}
|\Psi^+\rangle_{1\textit{x}} &=& \frac{|e\rangle_1|1\rangle_x+|g\rangle_1|0\rangle_x}{\sqrt{2}}
\nonumber \\
|\Psi^-\rangle_{1\textit{x}} &=& \frac{|e\rangle_1|1\rangle_x-|g\rangle_1|0\rangle_x}{\sqrt{2}}
\nonumber \\
|\Phi^-\rangle_{1\textit{x}} &=& \frac{|e\rangle_1|0\rangle_x-|g\rangle_1|1\rangle_x}{\sqrt{2}}
\nonumber \\
|\Phi^+\rangle_{1\textit{x}} &=& \frac{|e\rangle_1|0\rangle_x+|g\rangle_1|1\rangle_x}{\sqrt{2}}
\end{eqnarray}
Charlie performs Bell state projection between qubit-1 and qubit-\textit{x} (Bell state measurement (BSM)) and communicates his results to Bob which we have presented in Sec.~\ref{sec:bprojec} below.
%
%
\subsection{Bell state measurement}
\label{sec:bprojec}
BSM is realized at Charlie's end. Projection of a state $|\Lambda\rangle$ onto  $|\Sigma\rangle$ is defined as \cite{scherer2019mathematics}

\begin{equation}
P_\Sigma:=\langle{\Sigma}|\Lambda\rangle\hspace{1mm}|\Sigma\rangle
\label{eq:bellproj}
\end{equation}
Using $|\chi\rangle$ from Eq.~\eqref{eq:systems2} and applying Eq.~\eqref{eq:bellproj} we obtain a Bell state projection outcome communicated to Bob in the form
\begin{subequations}
\begin{equation}
_{1\textit{x}}\langle{\Psi^-}|\chi\rangle=\frac{1}{2}\left(\frac{|e\rangle_3|g\rangle_2-|g\rangle_3|e\rangle_2}{\sqrt{2}}\right)=\frac{1}{2}|\Psi^-\rangle_{32}
\label{eq:bellproj1}
\end{equation}
The Bell state $|\Psi^-\rangle_{32}$ in Eq.~\eqref{eq:bellproj1} is in the form of Alice's 
qubit in Eq.~\eqref{eq:twoatom}. Alice and Bob now 
have a Bell pair between qubit-2 and qubit-3. 
Similarly the other three Bell projections take the forms
\begin{equation}
_{1\textit{x}}\langle{\Psi^+}|\chi\rangle=\frac{1}{2}\left(\frac{|e\rangle_3|g\rangle_2+|g\rangle_3|e\rangle_2}{\sqrt{2}}\right)=\frac{1}{2}|\Psi^+\rangle_{32}
\label{eq:bellproj2}
\end{equation}
\begin{equation}
_{1\textit{x}}\langle{\Phi^-}|\chi\rangle=\frac{1}{2}\left(\frac{|e\rangle_3|e\rangle_2-|g\rangle_3|g\rangle_2}{\sqrt{2}}\right)=\frac{1}{2}|\Phi^-\rangle_{32}
\label{eq:bellproj3}
\end{equation}
\begin{equation}
_{1\textit{x}}\langle{\Phi^+}|\chi\rangle=-\frac{1}{2}\left(\frac{|e\rangle_3|e\rangle_2+|g\rangle_3|g\rangle_2}{\sqrt{2}}\right)=-\frac{1}{2}|\Phi^+\rangle_{32}
\label{eq:bellproj4}
\end{equation}
\end{subequations}
For these cases of Bell state projections in Eqs.~\eqref{eq:bellproj2}, \eqref{eq:bellproj3} and \eqref{eq:bellproj4}  it will be necessary for Bob to perform local corrections to qubit-3 by Pauli operators as shown in Tab.~\ref{tab:bobsaction}.  We also see that the probability of measuring states $|\psi\rangle_{32}$ in Eqs.~\eqref{eq:bellproj1}-\eqref{eq:bellproj4} in Charlie's lab is $p=\frac{1}{4}$. In general, by application of the entanglement swapping protocol (teleportation of entanglement), qubit-2 belonging to Alice and qubit-3 belonging to Bob despite never having interacted before became entangled. Further, we see that a maximally entangled anti-symmetric atom-field transition state $|\overline{\phi}_{g0}\rangle$ (in Eq.~\eqref{eq:transition6}) easily generated in the AJC interaction, can be used in quantum information processing (QIP) protocols like entanglement swapping (teleportation of entanglement) which we have demonstrated in this work. We note that it is not possible to generate such an entangled anti-symmetric state in the JC interaction starting with the atom initially in the ground state and the field mode in the vacuum state \cite{omolo2019photospins}. Recall that the JC interaction produces a meaningful physical effect, namely, spontaneous emission only when the atom is initially in the excited state and the field mode in the vacuum state.
\begin{center}
\begin{table*}
\caption{Table showing how Bob applies an appropriate gate to his qubit based on BSM from Charlie}
\label{tab:bobsaction}
\begin{tabular}{p{0.25 \linewidth}p{0.25 \linewidth}p{0.25\linewidth}}\\
\hline
\textbf{$|\varphi\rangle_{12}$}&
\textbf{$|\psi\rangle_{32}$} 
& UNITARY OPERATION\\
\hline\hline
$\frac{1}{\sqrt{2}}(|e\rangle_1|g\rangle_2-|g\rangle_1|e\rangle_2)$ 
& 
$\frac{1}{\sqrt{2}}(-|g\rangle_3|g\rangle_2+|e\rangle_3|e\rangle_2)$ 
& 
$-\hat{\sigma}_{x(atom3)}\otimes\hat{I}_{(atom2)}$
\\
& 
$\frac{1}{\sqrt{2}}(-|g\rangle_3|g\rangle_2-|e\rangle_3|e\rangle_2)$ 
& 
$-i\hat{\sigma}_{y(atom3)}\otimes\hat{I}_{(atom2)}$
\\
& 
$\frac{1}{\sqrt{2}}(|e\rangle_3|g\rangle_2+|g\rangle_3|e\rangle_2)$ 
& $\hat{\sigma}_{z(atom3)}\otimes\hat{I}_{(atom2)}$
\\
\hline
\end{tabular}
\end{table*}
\end{center}

%
%
\subsection{Maximal teleportation fidelity}

For any two-qubit state $\hat{\rho}$ the maximal fidelity is given by \citep{horodecki1999general}

\begin{equation}
F_{\hat{\rho}}=\frac{2f_{\hat{\rho}}+1}{3}
\label{eq:fidelity}
\end{equation}
where $f_{\hat{\rho}}$ is the fully entangled fraction defined in the form \cite{bennett1996mixed}

\begin{equation}
f_{\hat{\rho}}=\underset{|\Psi\rangle}{max}\langle\Psi|\hat{\rho}|\Psi\rangle=\left\lbrace{tr}\sqrt{\hat{\rho}_{expected}^{\frac{1}{2}}\hat{\rho}_{measured}\hat{\rho}_{expected}^{\frac{1}{2}}}\right\rbrace^2
\label{eq:fullyent}
\end{equation}
From Tab.~\ref{tab:bobsaction}

\begin{eqnarray}
\hat{\rho}_{expected} &=& |\varphi_{12}\rangle\langle\varphi_{12}|
\nonumber \\
&=& \frac{1}{2}\Big[(|e_1\rangle|g_2\rangle-|g_1\rangle|e_2\rangle)(\langle{e_1}|\langle{g_2}|-\langle{g_1}|\langle{e_2}|)\Big]
\nonumber \\
&=& \frac{1}{2}\Big[|e_1,g_2\rangle\langle{e_1},g_2|-|e_1,g_2\rangle\langle{g_1},e_2|\nonumber\\&-&|g_1,e_2\rangle\langle{e_1},g_2|+|g_1,e_2\rangle\langle{g_1},e_2|\Big]
\nonumber \\ 
&=& \frac{1}{2}
\begin{pmatrix}
0 & 0 & 0 & 0\\
0 & 1 & -1 & 0\\
0 & -1 & 1 & 0\\
0 & 0 & 0 & 0\\
\end{pmatrix}\nonumber\\&&
\label{eq:expected}
\end{eqnarray}

\begin{eqnarray}
\hat{\rho}_{measured} &=& |\psi_{32}\rangle\langle\psi_{32}|
\nonumber \\
&=& \frac{1}{2}\Big[(|e_3\rangle|g_2\rangle-|g_3\rangle|e_2\rangle)(\langle{e_3}|\langle{g_2}|-\langle{g_3}|\langle{e_2}|)\Big]
\nonumber \\
&=& 
\frac{1}{2}\Big[|e_3,g_2\rangle\langle{e_3},g_2|-|e_3,g_2\rangle\langle{g_3},e_2|\nonumber\\ &-&|g_3,e_2\rangle\langle{e_3},g_2|+|g_3,e_2\rangle\langle{g_3},e_2|
\Big]
\nonumber \\
&=&\frac{1}{2}
\begin{pmatrix}
0 & 0 & 0 & 0\\
0 & 1 & -1 & 0\\
0 & -1 & 1 & 0\\
0 & 0 & 0 & 0\\
\end{pmatrix}\nonumber\\&&
\label{eq:measured}
\end{eqnarray}
Substituting the results in Eq.~\eqref{eq:expected} and Eq.~\eqref{eq:measured} into the fully entangled fraction Eq.~\eqref{eq:fullyent} we obtain
\begin{equation}
f_{\hat{\rho}}=\left\lbrace{tr}
\begin{pmatrix}
0 & 0 & 0 & 0\\
0 & \frac{1}{2} & -\frac{1}{2} & 0\\
0 & -\frac{1}{2} & \frac{1}{2} & 0\\
0 & 0 & 0 & 0\\
\end{pmatrix}
\right\rbrace^2=1
\end{equation}
Substituting the value of the fully entangled fraction 
into Eq.~\eqref{eq:fidelity} we get
\begin{equation}
F_{\hat{\rho}}=1
\end{equation}
a maximal teleportation fidelity of unity, showing that the state was fully recovered, i.e Alice's qubit in Eq.~\eqref{eq:twoatom} was successfully teleported to 
Bob. We obtain an equal outcome to all the other 
measured states. We have thus achieved teleportation 
using a maximally entangled qubit state generated in an AJC interaction, using the case where the atom and field are initially in the absolute ground state $|g\rangle$, $|0\rangle$ as an example.
%
%
\section{ Conclusion}
\label{sec:conclusion}
In this paper we have analysed entanglement of a two-level atom and a quantized electromagnetic field mode in an AJC qubit formed in the AJC interaction mechanism. The effect of sum-frequency parameter and photon number on the dynamical behavior of entanglement measured by von Neumann entropy was studied which brought a clear visualization of this interaction similar to the graphical representation on Bloch sphere. The graphical representation of Rabi oscillations on the Bloch sphere demonstrated an important physical property, that the AJC interaction process occurs in the reverse sense relative to the JC interaction process. We further generated an entangled AJC qubit state in the AJC interaction mechanism which we used in the entanglement swapping protocol as Bob's qubit. We obtained an impressive maximal teleportation fidelity $F_\rho=1$ showing that the state was fully recovered. This impressive result of fidelity, opens all possible directions for future research in teleportation strictly within the AJC model. In conclusion we observe that the operator ordering that distinguishes the rotating (JC) component and anti-rotating component (AJC) has an important physical foundation with reference to the rotating positive and anti-rotating negative frequency components of the field mode which dictates the coupling of the degenerate states of a two-level atom to the frequency components of the field mode, an important basis for realizing the workings in the AJC interaction mechanism and JC interaction mechanism. 
%
%
\section*{Acknowledgment}
We thank Maseno University Department of Physics and Materials Science for providing a conducive environment to do this work.

%
%
 \bibliographystyle{apsrev}

\bibliography{../bib/cmayero}

\begin{thebibliography}{27}
\expandafter\ifx\csname natexlab\endcsname\relax\def\natexlab#1{#1}\fi
\expandafter\ifx\csname bibnamefont\endcsname\relax
  \def\bibnamefont#1{#1}\fi
\expandafter\ifx\csname bibfnamefont\endcsname\relax
  \def\bibfnamefont#1{#1}\fi
\expandafter\ifx\csname citenamefont\endcsname\relax
  \def\citenamefont#1{#1}\fi
\expandafter\ifx\csname url\endcsname\relax
  \def\url#1{\texttt{#1}}\fi
\expandafter\ifx\csname urlprefix\endcsname\relax\def\urlprefix{URL }\fi
\providecommand{\bibinfo}[2]{#2}
\providecommand{\eprint}[2][]{\url{#2}}

\bibitem[{\citenamefont{Braak}(2011)}]{braak2011integrability}
\bibinfo{author}{\bibfnamefont{D.}~\bibnamefont{Braak}},
  \bibinfo{journal}{Phys. Rev. Lett.} \textbf{\bibinfo{volume}{107}},
  \bibinfo{pages}{100401} (\bibinfo{year}{2011}).

\bibitem[{\citenamefont{Omolo}(2017{\natexlab{a}})}]{omolo2017conserved}
\bibinfo{author}{\bibfnamefont{J.~A.} \bibnamefont{Omolo}},
  \bibinfo{journal}{Preprint Research Gate, DOI:10.13140/RG.2.2.30936.80647}
  (\bibinfo{year}{2017}{\natexlab{a}}).

\bibitem[{\citenamefont{Omolo}(2017{\natexlab{b}})}]{omolo2017polariton}
\bibinfo{author}{\bibfnamefont{J.~A.} \bibnamefont{Omolo}},
  \bibinfo{journal}{preprint Research Gate, DOI:10.13140/RG.2.2.11833.67683}
  (\bibinfo{year}{2017}{\natexlab{b}}).

\bibitem[{\citenamefont{Omolo}(2019)}]{omolo2019photospins}
\bibinfo{author}{\bibfnamefont{J.~A.} \bibnamefont{Omolo}},
  \bibinfo{journal}{preprint Research Gate, DOI: 10.13140/RG.2.2.27331.96807}
  (\bibinfo{year}{2019}).

\bibitem[{\citenamefont{Born and Wolf}(1999)}]{born1999principles}
\bibinfo{author}{\bibfnamefont{M.}~\bibnamefont{Born}} \bibnamefont{and}
  \bibinfo{author}{\bibfnamefont{E.}~\bibnamefont{Wolf}},
  \emph{\bibinfo{title}{Principles of optics: electromagnetic theory of
  propagation, interference and diffraction of light, 7th edition}}
  (\bibinfo{publisher}{Cambridge university press}, \bibinfo{year}{1999}).

\bibitem[{\citenamefont{Boyer et~al.}(2017)\citenamefont{Boyer, Liss, and
  Mor}}]{boyer2017geometry}
\bibinfo{author}{\bibfnamefont{M.}~\bibnamefont{Boyer}},
  \bibinfo{author}{\bibfnamefont{R.}~\bibnamefont{Liss}}, \bibnamefont{and}
  \bibinfo{author}{\bibfnamefont{T.}~\bibnamefont{Mor}},
  \bibinfo{journal}{Phys. Rev. A} \textbf{\bibinfo{volume}{95}},
  \bibinfo{pages}{032308} (\bibinfo{year}{2017}).

\bibitem[{\citenamefont{Regula and Adesso}(2016)}]{regula2016entanglement}
\bibinfo{author}{\bibfnamefont{B.}~\bibnamefont{Regula}} \bibnamefont{and}
  \bibinfo{author}{\bibfnamefont{G.}~\bibnamefont{Adesso}},
  \bibinfo{journal}{Phys. Rev. Lett.} \textbf{\bibinfo{volume}{116}},
  \bibinfo{pages}{070504} (\bibinfo{year}{2016}).

\bibitem[{\citenamefont{Nielsen and Chuang}(2011)}]{nielsen2011quantum}
\bibinfo{author}{\bibfnamefont{M.~A.} \bibnamefont{Nielsen}} \bibnamefont{and}
  \bibinfo{author}{\bibfnamefont{I.~L.} \bibnamefont{Chuang}},
  \emph{\bibinfo{title}{Quantum Computation and Quantum Information}}
  (\bibinfo{publisher}{Cambridge University Press}, \bibinfo{year}{2011}).

\bibitem[{\citenamefont{Rieffel and Polak}(2011)}]{rieffel2011quantum}
\bibinfo{author}{\bibfnamefont{E.~G.} \bibnamefont{Rieffel}} \bibnamefont{and}
  \bibinfo{author}{\bibfnamefont{W.~H.} \bibnamefont{Polak}},
  \emph{\bibinfo{title}{Quantum computing: A gentle introduction}}
  (\bibinfo{publisher}{MIT Press}, \bibinfo{year}{2011}).

\bibitem[{\citenamefont{Van~Meter}(2014)}]{van2014quantum}
\bibinfo{author}{\bibfnamefont{R.}~\bibnamefont{Van~Meter}},
  \emph{\bibinfo{title}{Quantum networking}} (\bibinfo{publisher}{John Wiley \&
  Sons}, \bibinfo{year}{2014}).

\bibitem[{\citenamefont{Enr{\'\i}quez et~al.}(2014)\citenamefont{Enr{\'\i}quez,
  Reyes, and Rosas-Ortiz}}]{enriquez2014note}
\bibinfo{author}{\bibfnamefont{M.}~\bibnamefont{Enr{\'\i}quez}},
  \bibinfo{author}{\bibfnamefont{R.}~\bibnamefont{Reyes}}, \bibnamefont{and}
  \bibinfo{author}{\bibfnamefont{O.}~\bibnamefont{Rosas-Ortiz}},
  \bibinfo{journal}{J. Phys. Conf. Ser.} \textbf{\bibinfo{volume}{512}},
  \bibinfo{pages}{012021} (\bibinfo{year}{2014}).

\bibitem[{\citenamefont{Von~Neumann}(2018)}]{von2018mathematical}
\bibinfo{author}{\bibfnamefont{J.}~\bibnamefont{Von~Neumann}},
  \emph{\bibinfo{title}{Mathematical foundations of quantum mechanics: New
  edition}} (\bibinfo{publisher}{Princeton university press},
  \bibinfo{year}{2018}).

\bibitem[{\citenamefont{Wootters}(2001)}]{wootters2001entanglement}
\bibinfo{author}{\bibfnamefont{W.~K.} \bibnamefont{Wootters}},
  \bibinfo{journal}{Quantum Inf. Comput.} \textbf{\bibinfo{volume}{1}},
  \bibinfo{pages}{27} (\bibinfo{year}{2001}).

\bibitem[{\citenamefont{Abdel-Khalek}(2011)}]{abdel2011dynamics}
\bibinfo{author}{\bibfnamefont{S.}~\bibnamefont{Abdel-Khalek}},
  \bibinfo{journal}{J. Russ. Laser} \textbf{\bibinfo{volume}{32}},
  \bibinfo{pages}{86} (\bibinfo{year}{2011}).

\bibitem[{\citenamefont{Bennett
  et~al.}(1996{\natexlab{a}})\citenamefont{Bennett, DiVincenzo, Smolin, and
  Wootters}}]{bennett1996mixed}
\bibinfo{author}{\bibfnamefont{C.~H.} \bibnamefont{Bennett}},
  \bibinfo{author}{\bibfnamefont{D.~P.} \bibnamefont{DiVincenzo}},
  \bibinfo{author}{\bibfnamefont{J.~A.} \bibnamefont{Smolin}},
  \bibnamefont{and} \bibinfo{author}{\bibfnamefont{W.~K.}
  \bibnamefont{Wootters}}, \bibinfo{journal}{Phys. Rev. A}
  \textbf{\bibinfo{volume}{54}}, \bibinfo{pages}{3824}
  (\bibinfo{year}{1996}{\natexlab{a}}).

\bibitem[{\citenamefont{Bennett
  et~al.}(1996{\natexlab{b}})\citenamefont{Bennett, Bernstein, Popescu, and
  Schumacher}}]{bennett1996concentrating}
\bibinfo{author}{\bibfnamefont{C.~H.} \bibnamefont{Bennett}},
  \bibinfo{author}{\bibfnamefont{H.~J.} \bibnamefont{Bernstein}},
  \bibinfo{author}{\bibfnamefont{S.}~\bibnamefont{Popescu}}, \bibnamefont{and}
  \bibinfo{author}{\bibfnamefont{B.}~\bibnamefont{Schumacher}},
  \bibinfo{journal}{Phys. Rev. A} \textbf{\bibinfo{volume}{53}},
  \bibinfo{pages}{2046} (\bibinfo{year}{1996}{\natexlab{b}}).

\bibitem[{\citenamefont{Mohammadi and Jami}(2019)}]{mohammadi2019time}
\bibinfo{author}{\bibfnamefont{M.}~\bibnamefont{Mohammadi}} \bibnamefont{and}
  \bibinfo{author}{\bibfnamefont{S.}~\bibnamefont{Jami}},
  \bibinfo{journal}{Optik} \textbf{\bibinfo{volume}{181}}, \bibinfo{pages}{582}
  (\bibinfo{year}{2019}).

\bibitem[{\citenamefont{Liu et~al.}(2018)\citenamefont{Liu, Lu, Zhang, Liu, Li,
  Liang, Ma, Weng, Zhang, Liu et~al.}}]{liu2018nonlinear}
\bibinfo{author}{\bibfnamefont{X.-J.} \bibnamefont{Liu}},
  \bibinfo{author}{\bibfnamefont{J.-B.} \bibnamefont{Lu}},
  \bibinfo{author}{\bibfnamefont{S.-Q.} \bibnamefont{Zhang}},
  \bibinfo{author}{\bibfnamefont{J.-P.} \bibnamefont{Liu}},
  \bibinfo{author}{\bibfnamefont{H.}~\bibnamefont{Li}},
  \bibinfo{author}{\bibfnamefont{Y.}~\bibnamefont{Liang}},
  \bibinfo{author}{\bibfnamefont{J.}~\bibnamefont{Ma}},
  \bibinfo{author}{\bibfnamefont{Y.-J.} \bibnamefont{Weng}},
  \bibinfo{author}{\bibfnamefont{Q.-R.} \bibnamefont{Zhang}},
  \bibinfo{author}{\bibfnamefont{H.}~\bibnamefont{Liu}}, \bibnamefont{et~al.},
  \bibinfo{journal}{Int. J. Theor. Phys.} \textbf{\bibinfo{volume}{57}},
  \bibinfo{pages}{290} (\bibinfo{year}{2018}).

\bibitem[{\citenamefont{Zhang et~al.}(2018)\citenamefont{Zhang, Lu, Liu, Liang,
  Li, Ma, Liu, and Wu}}]{zhang2018quantum}
\bibinfo{author}{\bibfnamefont{S.-Q.} \bibnamefont{Zhang}},
  \bibinfo{author}{\bibfnamefont{J.-B.} \bibnamefont{Lu}},
  \bibinfo{author}{\bibfnamefont{X.-J.} \bibnamefont{Liu}},
  \bibinfo{author}{\bibfnamefont{Y.}~\bibnamefont{Liang}},
  \bibinfo{author}{\bibfnamefont{H.}~\bibnamefont{Li}},
  \bibinfo{author}{\bibfnamefont{J.}~\bibnamefont{Ma}},
  \bibinfo{author}{\bibfnamefont{J.-P.} \bibnamefont{Liu}}, \bibnamefont{and}
  \bibinfo{author}{\bibfnamefont{X.-Y.} \bibnamefont{Wu}},
  \bibinfo{journal}{Int. J. Theor. Phys.} \textbf{\bibinfo{volume}{57}},
  \bibinfo{pages}{279} (\bibinfo{year}{2018}).

\bibitem[{\citenamefont{Abdel-Khalek et~al.}(2015)\citenamefont{Abdel-Khalek,
  Quthami, and Ahmed}}]{abdel2015effects}
\bibinfo{author}{\bibfnamefont{S.}~\bibnamefont{Abdel-Khalek}},
  \bibinfo{author}{\bibfnamefont{M.}~\bibnamefont{Quthami}}, \bibnamefont{and}
  \bibinfo{author}{\bibfnamefont{M.}~\bibnamefont{Ahmed}},
  \bibinfo{journal}{Opt. Rev.} \textbf{\bibinfo{volume}{22}},
  \bibinfo{pages}{25} (\bibinfo{year}{2015}).

\bibitem[{\citenamefont{{\.Z}ukowski et~al.}(1993)\citenamefont{{\.Z}ukowski,
  Zeilinger, Horne, and Ekert}}]{zukowski1993event}
\bibinfo{author}{\bibfnamefont{M.}~\bibnamefont{{\.Z}ukowski}},
  \bibinfo{author}{\bibfnamefont{A.}~\bibnamefont{Zeilinger}},
  \bibinfo{author}{\bibfnamefont{M.~A.} \bibnamefont{Horne}}, \bibnamefont{and}
  \bibinfo{author}{\bibfnamefont{A.~K.} \bibnamefont{Ekert}},
  \bibinfo{journal}{Phys. Rev. Lett.} \textbf{\bibinfo{volume}{71}},
  \bibinfo{pages}{4287} (\bibinfo{year}{1993}).

\bibitem[{\citenamefont{Megidish et~al.}(2013)\citenamefont{Megidish, Halevy,
  Shacham, Dvir, Dovrat, and Eisenberg}}]{megidish2013entanglement}
\bibinfo{author}{\bibfnamefont{E.}~\bibnamefont{Megidish}},
  \bibinfo{author}{\bibfnamefont{A.}~\bibnamefont{Halevy}},
  \bibinfo{author}{\bibfnamefont{T.}~\bibnamefont{Shacham}},
  \bibinfo{author}{\bibfnamefont{T.}~\bibnamefont{Dvir}},
  \bibinfo{author}{\bibfnamefont{L.}~\bibnamefont{Dovrat}}, \bibnamefont{and}
  \bibinfo{author}{\bibfnamefont{H.}~\bibnamefont{Eisenberg}},
  \bibinfo{journal}{Phys. Rev. Lett.} \textbf{\bibinfo{volume}{110}},
  \bibinfo{pages}{210403} (\bibinfo{year}{2013}).

\bibitem[{\citenamefont{Pan et~al.}(1998)\citenamefont{Pan, Bouwmeester,
  Weinfurter, and Zeilinger}}]{pan1998experimental}
\bibinfo{author}{\bibfnamefont{J.-W.} \bibnamefont{Pan}},
  \bibinfo{author}{\bibfnamefont{D.}~\bibnamefont{Bouwmeester}},
  \bibinfo{author}{\bibfnamefont{H.}~\bibnamefont{Weinfurter}},
  \bibnamefont{and}
  \bibinfo{author}{\bibfnamefont{A.}~\bibnamefont{Zeilinger}},
  \bibinfo{journal}{Phys. Rev. Lett.} \textbf{\bibinfo{volume}{80}},
  \bibinfo{pages}{3891} (\bibinfo{year}{1998}).

\bibitem[{\citenamefont{Bose et~al.}(1998)\citenamefont{Bose, Vedral, and
  Knight}}]{bose1998multiparticle}
\bibinfo{author}{\bibfnamefont{S.}~\bibnamefont{Bose}},
  \bibinfo{author}{\bibfnamefont{V.}~\bibnamefont{Vedral}}, \bibnamefont{and}
  \bibinfo{author}{\bibfnamefont{P.~L.} \bibnamefont{Knight}},
  \bibinfo{journal}{Phys. Rev. A} \textbf{\bibinfo{volume}{57}},
  \bibinfo{pages}{822} (\bibinfo{year}{1998}).

\bibitem[{\citenamefont{Sousa and Roversi}(2019)}]{sousa2019selective}
\bibinfo{author}{\bibfnamefont{E.~H.} \bibnamefont{Sousa}} \bibnamefont{and}
  \bibinfo{author}{\bibfnamefont{J.}~\bibnamefont{Roversi}},
  \bibinfo{journal}{Quantum Rep.} \textbf{\bibinfo{volume}{1}},
  \bibinfo{pages}{63} (\bibinfo{year}{2019}).

\bibitem[{\citenamefont{Scherer}(2019)}]{scherer2019mathematics}
\bibinfo{author}{\bibfnamefont{W.}~\bibnamefont{Scherer}},
  \emph{\bibinfo{title}{Mathematics of Quantum Computing: An Introduction}}
  (\bibinfo{publisher}{Springer Nature}, \bibinfo{year}{2019}).

\bibitem[{\citenamefont{Horodecki et~al.}(1999)\citenamefont{Horodecki,
  Horodecki, and Horodecki}}]{horodecki1999general}
\bibinfo{author}{\bibfnamefont{M.}~\bibnamefont{Horodecki}},
  \bibinfo{author}{\bibfnamefont{P.}~\bibnamefont{Horodecki}},
  \bibnamefont{and}
  \bibinfo{author}{\bibfnamefont{R.}~\bibnamefont{Horodecki}},
  \bibinfo{journal}{Phys. Rev. A} \textbf{\bibinfo{volume}{60}},
  \bibinfo{pages}{1888} (\bibinfo{year}{1999}).

\end{thebibliography}

\end{document}